\documentclass[
    ,final            % use final for the camera ready runs
  ]
  {aipproc}

\layoutstyle{6x9}

\usepackage{graphicx}% Include figure files

\newcommand{\ima}{{\mbox{Im}\,}}

\begin{document}

\title{Chiral extrapolation of the $\sigma$ and $\rho$ mesons
  from dispersion relations and Chiral Perturbation Theory}


\author{G. R\'{i}os}{
  address={Dept. F\'{i}sica Te\'{o}rica II. 
    Universidad Complutense, 28040, Madrid. Spain}
}

\author{A. G\'{o}mez Nicola}{
  address={Dept. F\'{i}sica Te\'{o}rica II. 
    Universidad Complutense, 28040, Madrid. Spain}
}

\author{C. Hanhart}{address={Institut f\"{u}r Kernphysik (Theorie), Forschungzentrum
  J\"{u}lich, D-52425 J\"{u}lich, Germany}
}

\author{J. R. P\'{e}laez}{
  address={Dept. F\'{i}sica Te\'{o}rica II. 
    Universidad Complutense, 28040, Madrid. Spain}
}

\begin{abstract}
We review our recent study of the pion mass 
dependence of the $\rho$ and $\sigma$ 
resonances, generated from
one-loop $SU(2)$ Chiral Perturbation Theory (ChPT)
with the Inverse Amplitude Method (IAM). 
In order to properly account
for the Adler zero region, we also review the recently
obtained modified version of the IAM; which is based on analyticity, 
elastic unitarity and ChPT 
at low energies, thus yielding the correct pion mass dependence of
the resonance pole positions up to next--to--leading order in ChPT. 
As main
results we find that the $\rho\pi\pi$ coupling constant is almost $m_\pi$
independent and that $M_\rho$ shows a smooth $m_\pi$ dependence
while that of the $\sigma$ shows a strong non-analyticity. These findings
are important for studies of the meson spectrum on the lattice.
\end{abstract}

\maketitle

\section{Introduction}
Light hadron spectroscopy lies beyond the realm
of perturbative QCD.
Although lattice QCD provides, in principle,
a rigorous way to extract non--perturbative quantities,
present calculations use relatively high quark masses, at least
for studies of scalar mesons, i. e.,
\cite{Aoki:1999ff,scalars}. Thus,  
appropriate extrapolation formulas are called for. 
Chiral Perturbation Theory (ChPT) \cite{chpt1} 
provides such extrapolations, since
it is built
as an expansion in momenta and masses,
generically $O(p/4\pi f_\pi)^2$, of a Lagrangian  
involving the Goldstone Bosons of the QCD chiral
symmetry breaking (pions), compatible
with all QCD symmetries.  
ChPT is renormalized order by order by absorbing loop
divergences in the parameters of higher order
counterterms (low energy constants - LECs), which 
are the coefficients of the energy and mass expansion, so
{\it they have no quark mass dependence}.
Their  values depend
on the QCD dynamics, and have to be determined from
experiment.
In SU(2)-ChPT
$\pi\pi$ scattering only four LECs appear, denoted $l_i$.

The ChPT expansion
provides a {\it systematic and model independent} 
description of how the observables depend on QCD
parameters, like
the quark masses, and this can be 
implemented systematically up to the desired 
order in the ChPT expansion.

We review here our recent derivation of a modified version
of the IAM \cite{modIAM}; based 
on dispersion theory, unitarity and ChPT to
next--to--leading order (NLO), 
which we use
to predict
the quark mass dependence of the
$\sigma$ and $\rho$ mesons \cite{Hanhart:2008mx}.

We focus only on the $\sigma$ and the $\rho$, so it is
enough to work with the lightest quarks 
$u,d$ in the isospin limit with a mass 
$\hat m=(m_u+m_d)/2$. Since $m_\pi$ is given by 
$m_\pi^2\sim \hat m+...$ \cite{chpt1},
studying the $\hat m$ dependence is equivalent to study
the $m_\pi$ dependence.

\section{Unitarization and Dispersion Theory}

The $\sigma$ and $\rho$ resonances appear as poles
in the second Riemann sheet of the $(I,J)=(0,0)$ and $(1,1)$ partial
waves of isospin $I$ and angular momentum $J$ respectively. 
For these partial waves, elastic unitarity implies, for physical values
of $s$:
\begin{equation}
  \label{unit}
{\rm Im }\: t(s)=\sigma (s)\vert t(s)\vert^2 \Rightarrow 
{\rm Im }\,\left( t(s)^{-1}\right)=-\sigma(s), \qquad {\rm with} \quad
\sigma(s)=2 p/\sqrt{s},
\end{equation}
where $s$ is the Mandelstamm variable and $p$ is the 
center of mass momentum. Consequently,
the imaginary part of the inverse amplitude is known exactly.
ChPT amplitudes, being an expansion $t=t_2+t_4+\cdots$ with
$t_k=O(p^k)$, satisfy Eq. \eqref{unit}
just perturbatively:
\begin{equation}
  \ima\, t_4(s)=\sigma(s)\vert t_2(s)\vert^2, 
  \quad\Rightarrow\quad	\ima\: {t_4(s)}/{t_2(s)^2}=\sigma(s),
  \label{pertunit}
\end{equation}
and cannot generate poles. Therefore the resonance region lies beyond
the reach of standard ChPT.  However, it can be reached by combining
ChPT with dispersion theory either for the amplitude~\cite{gilberto}
or the inverse amplitude through the
IAM~\cite{Truong:1988zp,Dobado:1996ps,Guerrero:1998ei}.

The elastic IAM \cite{modIAM} uses the 
ChPT series and elastic unitarity
to evaluate a dispersion relation for the inverse amplitude.
The analytic structure of $1/t$ consists on a right cut
from threshold to $\infty$, a left cut from $-\infty$ to
$s=0$, and possible poles coming from zeros of $t$.  
For scalar waves, $t$ vanishes at the so called Adler zero,
$s_A$, that lies on the real axis below threshold, thus within the
ChPT region of applicability. Its position can be obtained from the 
ChPT series, i.e., $s_A=s_2+s_4+\cdots$, where $t_2$ vanishes at $s_2$,
$t_2+t_4$ at $s_2+s_4$, etc. 

We write then a once subtracted dispersion relation for the inverse
amplitude, where we have chosen the subtraction point to be the
Adler zero:
\begin{equation}
  \label{eq:1/tdisp}
  \frac1{t(s)}=\frac{s-s_A}{\pi}
  \int_{RC}ds'\frac{{\rm Im}\,1/t(s')}{(s'-s_A)(s'-s)}+LC(1/t)+PC(1/t),
\end{equation}
where ``LC'' stands for a similar integral over the left cut and ``PC''
stands for the contribution of the pole at the Adler zero. 
Since $t_2$ is real
on the real axis and $t_4$ has the same analytic structure as $t$,we can
similarly write
\begin{equation}
  \label{t4/t2disp}
  \frac{t_4(s)}{t_2(s)^2}=\frac{s-s_2}{\pi}\int_{RC}ds'
  \frac{{\rm Im}\,t_4(s')/t_2(s')^2}{(s'-s_2)(s'-s)}
  +LC(t_4/t_2^2)+PC(t_4/t_2^2),
\end{equation}
where we have now subtracted at $s_2$, which is
the LO approximation to the Adler zero, 
and $PC$ stands for the contribution of the pole at $s_2$.
We can now use unitarity, Eqs. \eqref{unit} and \eqref{pertunit},
to find that the imaginary parts on the right cut of both dispersion
relations are {\it exactly opposite to each other}. 
Since the LC integral is weighted
at low energies, we can use ChPT to approximate
$LC(1/t)\simeq -LC(t_4/t_2^2)$. 
The pole contribution $PC(1/t)$ can also be
evaluated with ChPT since it involves 
derivatives of $t$ evaluated at the
Adler zero, where ChPT is perfectly justified. 
Finally, we approximate with
ChPT $(s-s_A)(s'-s_A)\simeq (s-s_2)/(s'-s_2)$. 
Altogether, we find a modified IAM (mIAM) formula:
\begin{equation}
  \label{mIAM}
  t^{mIAM}= \frac{t^2_2}{ t_2-t_4 +A^{mIAM}}\,,\quad
  A^{mIAM}=t_4(s_2){-}\frac{(s_2{-}s_A)(s{-}s_2)
    \left[t'_2(s_2){-}t'_4(s_2)\right]}{s{-}s_A}.
\end{equation}
The standard IAM is recovered for $A^{mIAM}=0$, which holds exactly for
all partial waves except the scalar ones. In the original IAM
derivation \cite{Truong:1988zp,Dobado:1996ps} $A^{mIAM}$ was neglected,
since it formally yields a NNLO contribution and  is numerically
very small, except near the Adler zero, where it diverges. However,
if $A^{mIAM}$ is neglected, the IAM Adler zero occurs at $s_2$,
correct only to LO, it is a double zero instead of a simple one, and
a spurious pole of the amplitude
appears close to the Adler zero. All of these caveats are removed
with the mIAM, Eq. \eqref{mIAM}. The differences in the 
physical and resonance region between the IAM and the mIAM are less
than 1\%. However, as we will see, for large $m_\pi$ the $\sigma$ pole
splits in two virtual poles below threshold, one of them moving towards
zero and approaching the Adler zero region, where the IAM fails. Thus, we will
use for our calculations the mIAM, although it is only relevant for the 
mentioned second $\sigma$ pole, and only when 
it is very close to the Adler zero.

\section{Results}

By changing $m_\pi$ in the amplitudes we see how the poles generated
with the IAM evolve.
We will use the LECs values 
$10^3l^r_3=0.8\pm 3.8$ and $10^3l^r_4=6.2\pm 5.7$ from \cite{chpt1}
and fit the mIAM to data up to the resonance region to find
$10^3l^r_1=-3.7\pm 0.2$ and $10^3l^r_2=5.0\pm 0.4$. These LECs 
are evaluated at $\mu=770$ MeV.

The values of $m_\pi$ considered should fall within the ChPT
range of applicability and allow for some elastic $\pi\pi$
regime below $K\bar K$ threshold. Both criteria are
satisfied if $m_\pi\leq 500$ MeV, since $SU(3)$ ChPT
still works with such kaon masses, and because for
$m_\pi\simeq 500$ MeV, the kaon mass becomes $\simeq 600$, leaving
200 MeV of elastic region.

\begin{figure}[t]
%\begin{center}
  \hbox{
    \includegraphics[scale=0.25,angle=0]{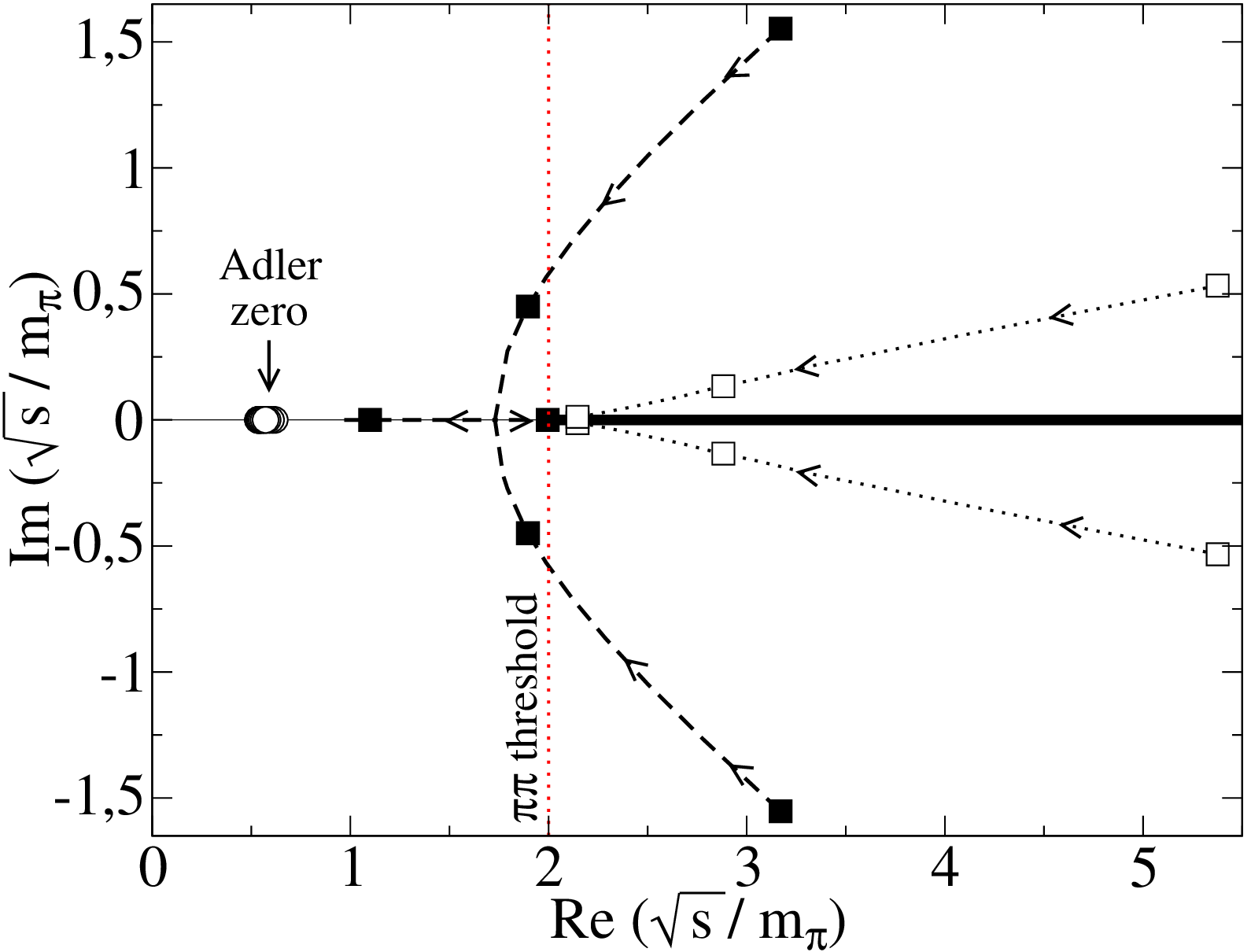}
    \includegraphics[scale=0.25,angle=0]{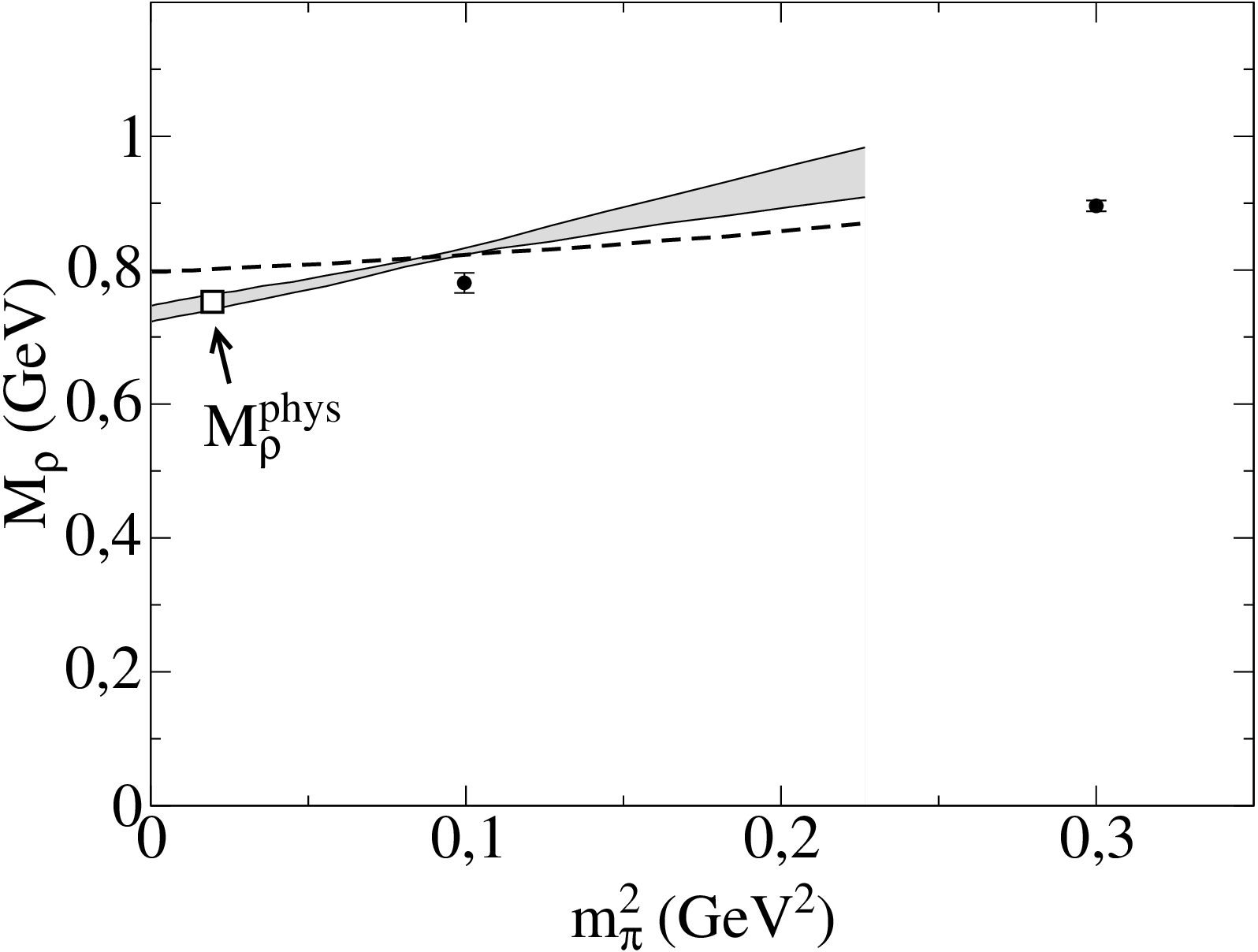}
  }
  \caption{ {\bf Left:} Movement of the $\sigma$ (dashed lines) 
    and $\rho$ (dotted
    lines) poles for increasing $m_\pi$ (direction indicated by the
    arrows) on the second sheet.  The filled (open) boxes denote the
    pole positions for the $\sigma$ ($\rho$) at pion masses $m_\pi=1,\
    2,$ and $3 \times m_\pi^{\rm phys}$, respectively. For
    $m_\pi=3m_\pi^{\rm phys}$ three poles accumulate in the plot 
    very near the $\pi\pi$ threshold. Note that all poles are
    always far from the Adler zero (circles).
    {\bf Right:} The grey band shows the $m_\pi$ dependence of $\rho$ pole
    mass from the IAM versus recent lattice results from
    \cite{Aoki:1999ff}. The dashed line is the IAM result for
    $N_c=10$.}
  \label{poles}
  % \end{center}
\end{figure}

Fig. \ref{poles} (left) shows the evolution of the $\sigma$ and $\rho$
pole positions as $m_\pi$ is increased. In order to see the pole
movements relative to the two pion threshold, which is also increasing,
all quantities are given in units of $m_\pi$, so the threshold is
fixed at $\sqrt{s}=2$. Both poles moves closer to threshold and
they approach the real axis. The $\rho$ poles reach the real axis
as the same time that they cross threshold.
One of them jumps into the first sheet and stays below
threshold in the real axis as a bound state, while its conjugate
partner remains on the second sheet practically at the very same
position as the one in the first. In contrast, the $\sigma$
poles go below threshold with a finite imaginary part before they
meet in the real axis, still on the second sheet, becoming
virtual states. As $m_\pi$ is increased further, one of the poles
moves toward threshold and jumps through the branch point to the
first sheet and stays in the real axis below threshold, very
close to it as $m_\pi$ keeps growing. The other $\sigma$ pole moves
down in energies further from threshold and remains 
on the second sheet.
This analytic structure, with
two very asymmetric poles in different sheets for a scalar wave, could be
a signal of a prominent molecular component \cite{Weinberg,baru}.
Similar pole movements have been also found within quark models
\cite{vanBeveren:2002gy}.

Note that ChPT has ben used to evaluate the subtraction constants
at the Adler zero and the low energy part of LC, 
{\it always far from the resonance poles}, 
even when they move below threshold, 
as shwon in Fig. \ref{poles} (left),
being the use of ChPT perfectly justified.

\begin{figure}
  \label{massandwidth}
  \centering
  \hbox{
    \includegraphics[scale=0.25,angle=0]{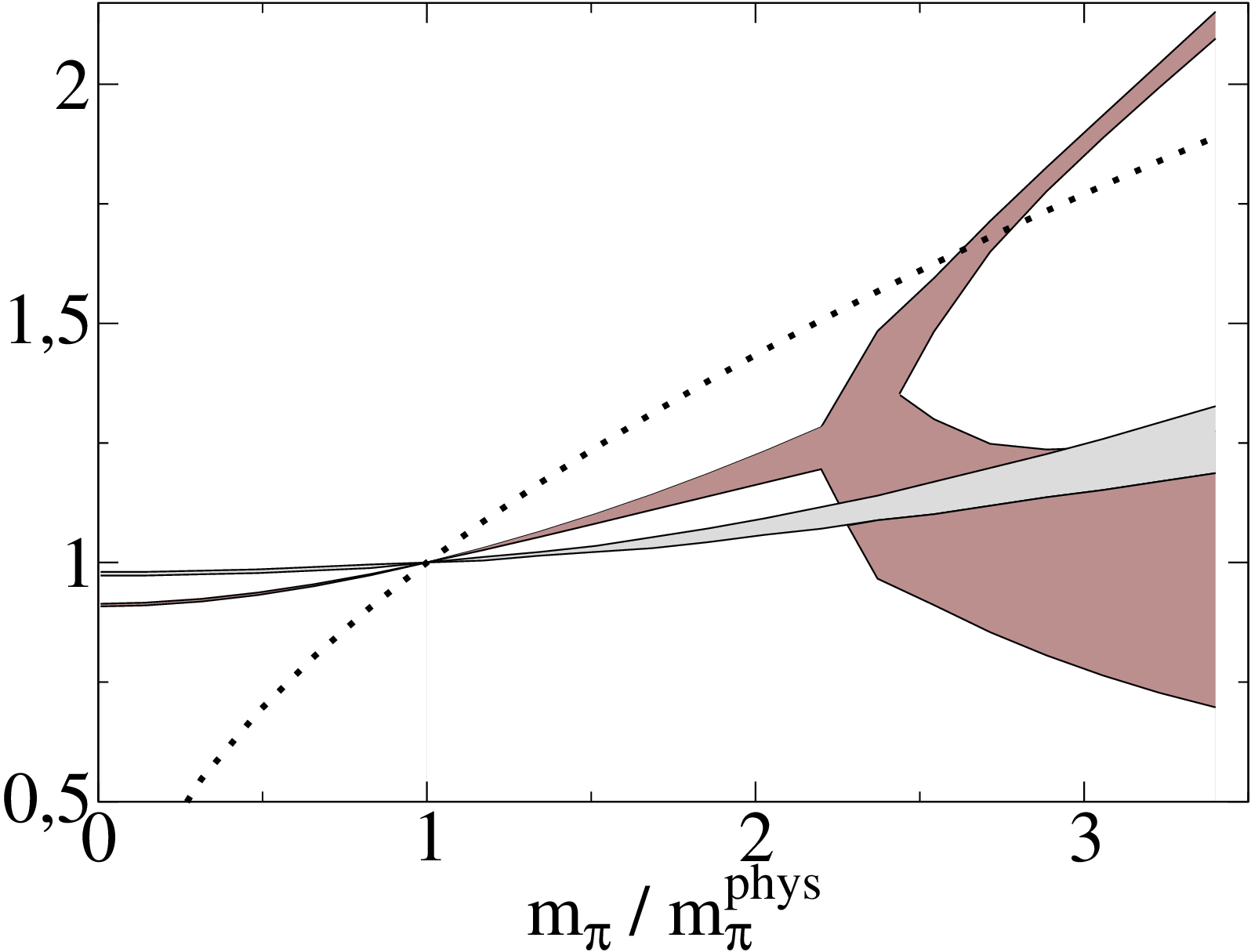}
    \includegraphics[scale=0.25,angle=0]{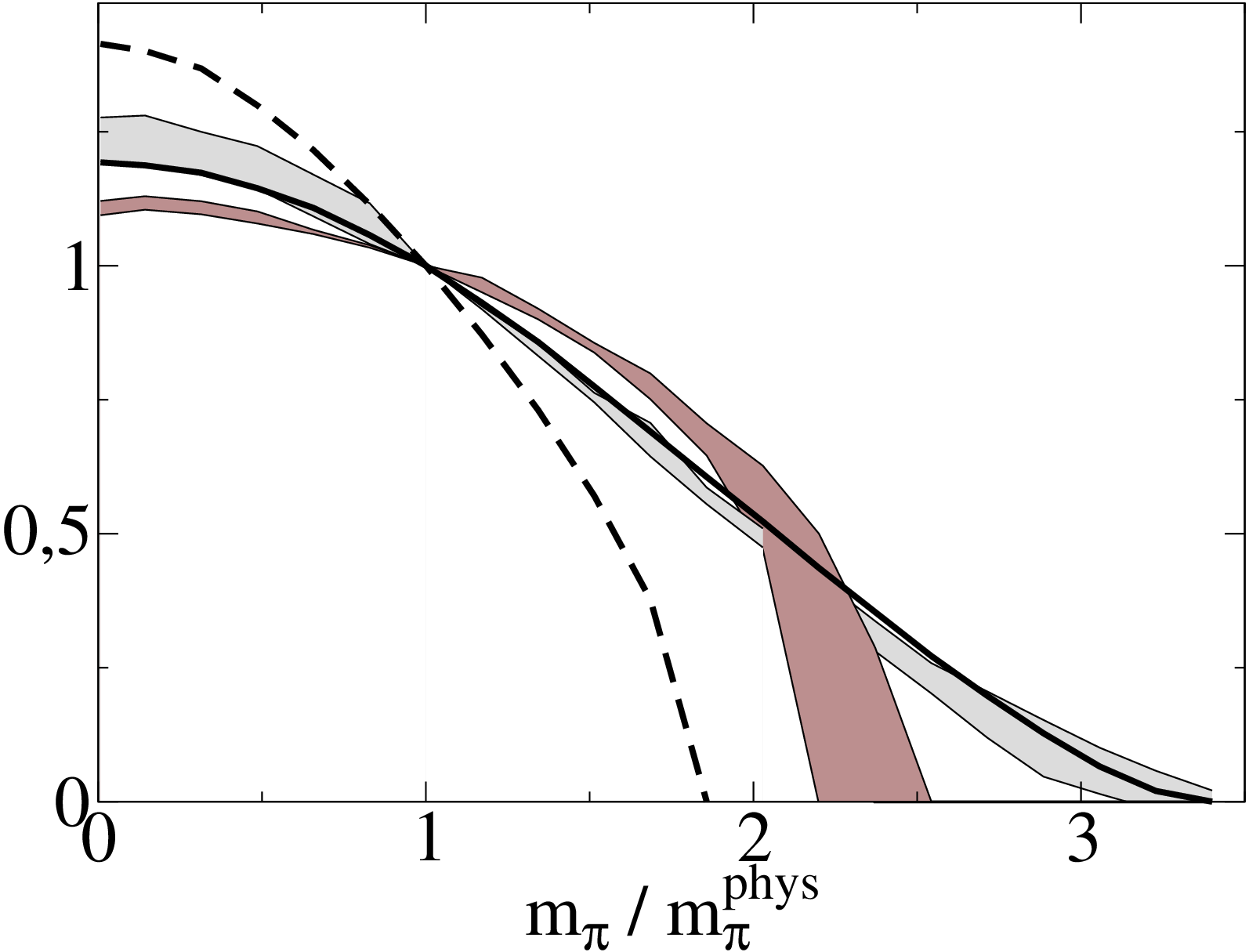}
  }

  \caption{ $m_\pi$ dependence of resonance masses (left)
  and widths (right) in units of the physical values. 
 In both panels the dark (light) band shows the results for
the $\sigma$ ($\rho$). The width of the bands reflects the uncertainties
induced from the uncertainties in the LECs. The dotted line 
shows the $\sigma$ mass dependence
  estimated in Ref.~\cite{Jeltema:1999na}. The dashed (continuous)
line shows the $m_\pi$ dependence of the $\sigma$ ($\rho$) 
width from the change of phase space only, assuming
a constant coupling of the resonance to $\pi\pi$.}
  
\end{figure}

In Fig. \ref{massandwidth} (left) we show the $m_\pi$ dependence
of $M_\rho$ and $M_\sigma$ (defined from the pole position
$\sqrt{s_{pole}}=M-i\Gamma /2$), normalized to their physical values.
The bands cover the LECs uncertainties. We see that both masses
grow with $m_\pi$, but $M_\sigma$ grows faster than $M_\rho$. 
Below $m_\pi\simeq 330$  MeV we only show one line because the two
conjugate $\sigma$ poles have the same mass. Above 330 MeV, these
two poles lie on the real axis with two different masses. The
heavier pole goes towards threshold and around $m_\pi\simeq 465$
moves into the
first sheet. Note also that the
$m_\pi$ dependence of $M_\sigma$ is much softer than suggested in
\cite{Jeltema:1999na}, shown as the dotted line, which in addition
does not show the two virtual poles.

In the right panel of Fig. \ref{massandwidth} we show the $m_\pi$
dependence of $\Gamma_\rho$ and $\Gamma_\sigma$ normalized to their
physical values, where we see that both widths become smaller. 
We compare this decrease with the expected reduction from
phase space as the resonances approach the $\pi\pi$ threshold.
We find that $\Gamma_\rho$ follows very well
this expected behavior, which
implies that the $\rho\pi\pi$ coupling is almost $m_\pi$ independent.
In contrast $\Gamma_\sigma$ shows a different behavior from the
phase space reduction expectation. This suggest a strong $m_\pi$
dependence of the $\sigma$ coupling to two pions, necessarily
present for molecular states \cite{baru,mol}.

Fig. \ref{poles} (right) is a comparison of our results for $M_\rho$
with some lattice results \cite{Aoki:1999ff}, which deserves
several words of caution. Our $M_\rho$ is the ``pole mass'',
which is deep in the complex plane, and, due to the 
momentum discretization induced by the
finite lattice volume, the minimum energy with which pions are
produced is larger than the resulting $M_\rho$, so the lattice
rho has no width. We can mimic in our formalism a narrower
$\rho$ by increasing the number of colors, $N_c$ \cite{Pelaez:2006nj}.
We also show the result for the rho mass for $N_c=10$.
We see that making the $\rho$ artificially narrower yields a 
better agreement with lattice data. With these caveats in mind
our results are in qualitative agreement with those of the lattice.
Following \cite{Bruns:2004tj} one may write
$M_\rho=M_\rho^0+c_1+O(m_\pi^3)$, where the $c_i$ parameters are
expected to be of order one and $M_\rho\sim 0.65-0.80$ GeV, which
is confirmed by a fit to lattice data \cite{Bruns:2004tj}.
We can fit our results and predict $M_\rho^0=0.735\pm 0.0017$ MeV
and $c_1=0.90\pm 0.17\: {\rm GeV}^{-1}$. Although the
pion mass dependence of our calculation is steeper than
that of the lattice, the values obtained are still consistent
with the expectations mentioned above. Let us remind that the
$m_\pi$ dependence in our approach is correct only up to
NLO in ChPT. 

We thank the organicers for creating the nice scientific atmosphere of
the workshop and the Spanish research contracts No. PR27/05-13955-BSCH,
No. FPA2007-29115-E, No. FPA2004-02602, No. UCM-CAM 910309 and No.
BFM2003-00856 
for partial finantial support.

\bibliographystyle{aipproc}   % if natbib is available

\end{document}